\documentclass[doublecol]{epl2}

\title{Estimating the principal components of correlation matrices from all their empirical eigenvectors}

\author{R\'emi Monasson\inst{1} \and Dario Villamaina\inst{1,2}}
\shortauthor{R. Monasson \etal}

\institute{                    
  \inst{1} Laboratoire de Physique Th\'eorique de l'Ecole Normale Sup\'erieure, associ\'e au CNRS et \`a l'Universit\'e Pierre et Marie Curie, 24 rue Lhomond 75005 Paris - France\\
  \inst{2}  Institut de Physique Th\'eorique Philippe Meyer, 24 rue Lhomond 75005 Paris - France
}

\abstract{ We consider the problem of estimating the principal
  components of a population covariance matrix from a limited number
  of measurement data. Using a combination of random matrix and
  information-theoretic tools, we show that all the eigenmodes of the
  sample correlation matrices are informative, and not only the top
  ones. We show how this information can be exploited when {\em prior}
  information about the principal component, such as whether it is
  localized or not, is available by mapping the estimation problem
  onto the search for the ground state of a spin-glass-like effective
  Hamiltonian encoding the prior. Results are illustrated numerically
  on the spiked covariance model.  }

\begin{document}

\maketitle

\section{\bf Introduction}

The availability of large-scale measurements of complex systems, such
as in biology, finance, sociology, ... calls for new methods to
extract information from those data. Of crucial importance is the
characterization of the correlation structure of the data, which
reflects the underlying interaction network between the system
components. A widely-used technique is principal component analysis
(PCA), which retains only the components corresponding to the largest
eigenvalues of the empirical correlation matrix computed from the
data, considered as most informative. PCA applications range from
computer vision~\cite{DB01} to finance~\cite{LCBP99}, to
neuroscience~\cite{CB14} and many others. PCA can, however, be
inefficient in some cases \cite{YR01}, in particular when the number
$T$ of available data is comparable to the number $N$ of system
components, a situation referred to as {\em high-dimensional} data
analysis \cite{D00}. 

 In this Letter we focus
on one aspect of this question, namely, how to estimate the main
eigenmode(s) of the `true' system correlation matrix at large $N/T$
ratio. We show that considering only the main components of the
empirical correlation matrix, as PCA does, is generally not optimal,
and that taking into account the eigenmodes associated to the low
eigenvalues can greatly improved the quality of the predictions.

\begin{figure}[h]
\begin{center}
\includegraphics[width=0.4\columnwidth,clip=true]{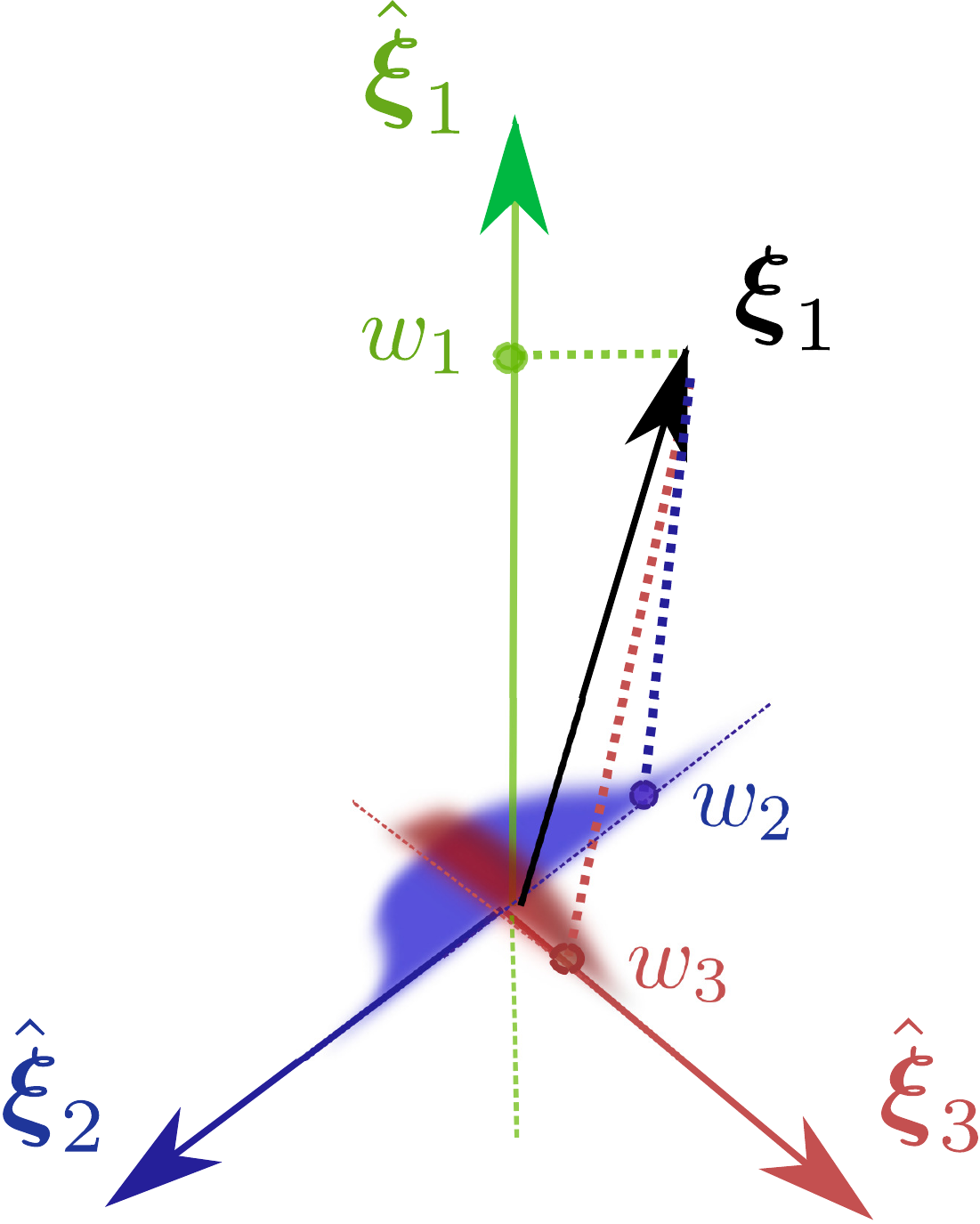}
\caption{Eigenvectors $\hat{\boldsymbol\xi}_m$ of the sample correlation matrix  and  `true' component ${\boldsymbol\xi}_1$  in dimension $N=3$. If the statistics is sufficient, {\em i.e.} $r$ is low enough, the overlap $w_1={\boldsymbol{\xi}}_1\cdot\hat{\boldsymbol{\xi}}_1$ is finite in the large $N$ limit, while the overlaps $w_m = {\boldsymbol{\xi}}_1\cdot\hat{\boldsymbol{\xi}}_{m}$ (with $m\ge 2$) vanish as $O(1/\sqrt N)$.  }
\end{center}
\end{figure}

To fix notations let us consider a collection of $N$ Gaussian random
variables $x_i$ ($i=1,\ldots , N$), with zero means and (population)
covariances $C_{ij}$. Assume we have observed $T$ independent
realizations of those variables, which define the $N\times
T$--dimensional rectangular matrix $X$, {\em e.g.} $X_{3,2}$
corresponds to the second observation of variable $x_{3}$.  The
empirical covariance matrix, $\hat{C}\equiv \frac 1T X\cdot
X^{\dagger}$, also called sample covariance matrix is an estimator of
the population covariance matrix $C$. Let us call $\hat
{\boldsymbol{\xi}}_{m}$, $m=1,2,..., N$, the normalized eigenvectors
of $\hat C$, corresponding to eigenvalues $\hat \lambda_1\ge \hat
\lambda_2\ge ... \ge \hat \lambda _m$. Similarly we call
${\boldsymbol{\xi}}_{m}$ and $\lambda_m$ the normalized eigenvectors
and eigenvalues (ranked in decreasing order) of $C$. Suppose now
  that $N$ is kept fixed and the number of observations $T$ is
  increased. A perfect sampling condition corresponds to the limit of infinite measurements ($T\to\infty$), where the matrix
  estimator $\hat{C}$ approaches to the true covariance $C$. This
  scenario changes if the number of variables $N$ increases at the same pace as the    number of measures $T$, with a fixed ratio $r=N/T$, hereafter called sampling
noise. In that case, the estimator $\hat{C}$ differs from the true covariance matrix since it is affected by finite sampling effects, a common situation in experiments. Perfect sampling is recovered in the limit case $r\to 0$.

While the typical \cite{S95} properties of the distribution of the eigenvalues of $\hat C$ are well characterized (and theoretical results for rare events are available in the case of purely uncorrelated variables \cite{DM06,MV12}), much less is known about its eigenvectors, see~\cite{HR04bis} or~\cite{AB12} and references therein. We want to estimate the top component $\boldsymbol{\xi}_1$. A simple estimator is provided by $\hat{\boldsymbol{\xi}}_1$, which is naturally expected to be exact when $r\to 0$ (in the absence of eigenvalue degeneracy), see Fig.~1. For finite $r$, however, $\hat{\boldsymbol\xi}_1$ is generally not a perfect estimator, and we show below how the knowledge of the other empirical eigenvectors $\hat{\boldsymbol{\xi}}_m$, with $m\ge 2$, may considerably help to improve the estimate of $\boldsymbol{\xi}_1$.

Let us consider the scalar products $w_m \equiv \boldsymbol{\xi}_1\cdot \hat{\boldsymbol{\xi}}_m$ (Fig.~1). Those overlaps are stochastic variables with zero mean, and variances $\left<w_m^2\right>$, where $\langle\cdot\rangle$ denotes the average over $X$. Each eigenvector $\hat{\boldsymbol{\xi}}_m$ taken individually is very weakly informative about $\boldsymbol{\xi}_1$ as the overlap vanishes as $N^{-1/2}$. On the contrary we show below that the mutual information between an extensive (of the order of $N$) number of eigenvectors $\hat{\boldsymbol{\xi}}_m (m\ge 2)$ and $\boldsymbol{\xi}_1$ remains finite when $N\to \infty$. We then present one application where the knowledge of the overlaps $w_m$ helps us to improve our prediction of the top component $\boldsymbol{\xi}_1$ in the presence of prior information about this vector (here, that it has ``large'' components).

\section{The spiked covariance model}

We will illustrate our approach on the spiked covariance model, a popular model in random matrix theory, in which all the eigenvalues of $C$, but one, say,
$\lambda_1\equiv \gamma$, are equal to unity. Eigenvalue $\gamma$ represents the `signal', with its associated eigenvector $\boldsymbol{\xi}_1$. We consider the case $\gamma>1$ below, but similar results are found for $\gamma<1$. Let $\hat\rho(\hat\lambda)=\displaystyle{\frac{1}{N}\sum_{m}\langle\delta(\hat\lambda-\lambda_{m})\rangle}$ be the average density of eigenvalues of $\hat C$ and $\rho(\lambda)$ the density of eigenvalues of $C$. 

For `weak' signals, $\gamma<\gamma_{c}(r)\equiv 1+\sqrt{r}$, $\hat \rho$ coincides with the spectrum of the covariance matrix of $N$ independent variables, the so-called Marchenko-Pastur (MP) distribution~\cite{MP67}, defined as
\begin{equation}
\hat\rho_{_{MP}}(\hat\lambda)=\frac{\sqrt{(\hat\lambda_{+}-\hat\lambda)(\hat\lambda-\hat\lambda_{-})}}{2\pi\hat\lambda r}\label{MPdistr}
\end{equation}
 where $\hat\lambda_{\pm}(r)=(1\pm\sqrt{r})^{2}$ are the edges of the distribution~\footnote{We consider here the case $r<1$. For $r>1$ a  $\delta$-peak in $\hat\lambda=0$ of mass $1-\frac{1}{r}$ is present.}. For `strong' signal, $\gamma>\gamma_{c}(r)$, the spectrum $\hat \rho$ is equal to the MP spectrum, and includes one eigenvalue, isolated from the MP bulk and centered in
\begin{equation}
\hat{\gamma} (\gamma)=\gamma+r\;\frac{\gamma}{\gamma-1} \ .\label{eq:signal_shift}
\end{equation}
The onset of a signal-related eigenvalue $\hat \gamma$ at the critical value of $\gamma=\gamma_c(r)$ was first reported in \cite{HR04} and mathematically proven in \cite{BBP05}. Similar `retarded learning' transitions are encountered in models of neural networks~\cite{WN94} and in the Gaussian Matrix ensemble~\cite{EJ76}. The transition is pictorially represented in Fig.~2.

\begin{figure}[h]
\includegraphics[width=1.\columnwidth]{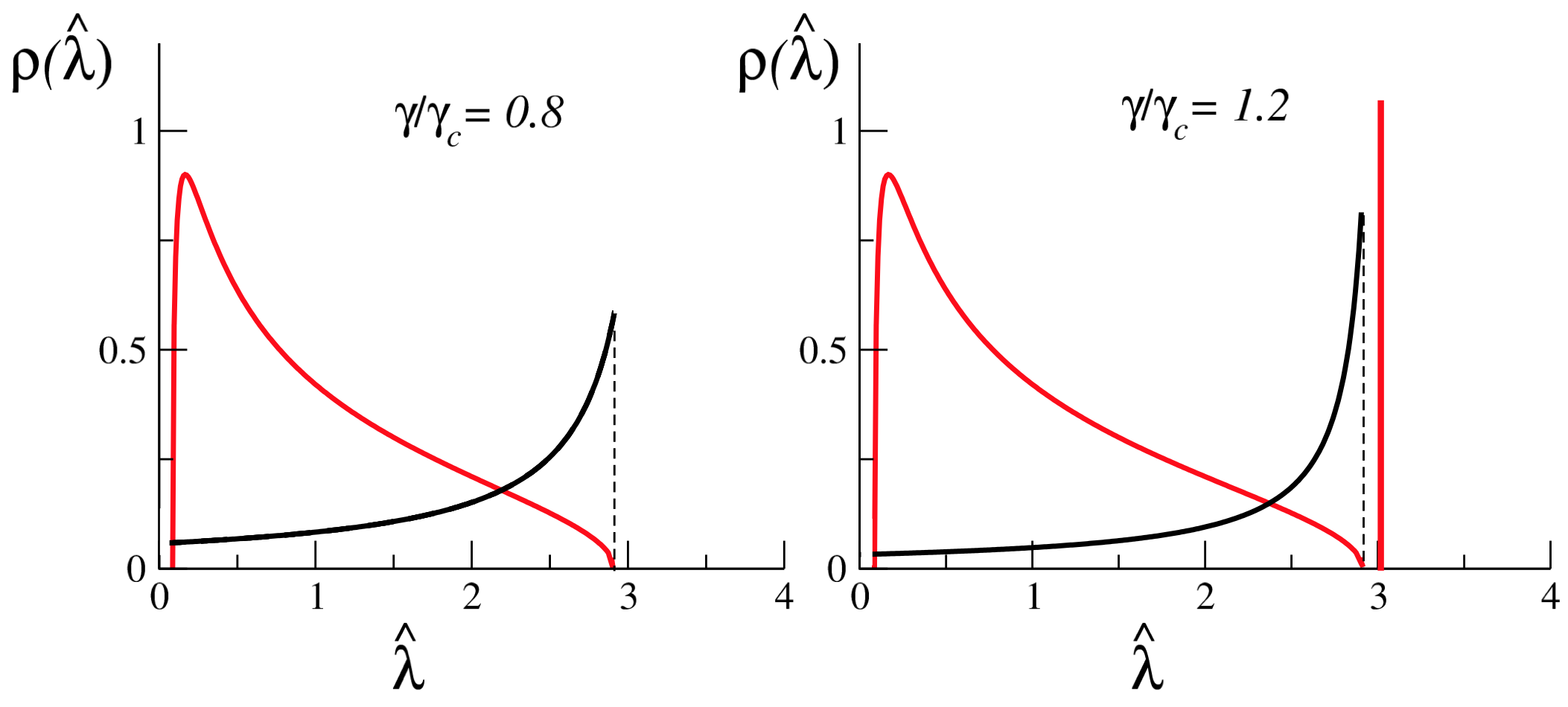}
\caption{The spiked covariance model below (left, $\gamma<\gamma_c(r)$) and above (right, $\gamma>\gamma_c(r)$). The eigenvalue
spectrum given by the Marchenko-Pastur distribution in Eq.~(\ref{MPdistr}) is shown in red; for $\gamma>\gamma_c(r)$, the signal eigenvalue  $\hat\gamma(\gamma)$, see Eq.~(\ref{eq:signal_shift}), is represented by a vertical red line. 
The squared overlap function $W^{2}(\gamma,\hat\lambda)$ in Eq.~(\ref{w_curve}) is shown in black over the interval
$[\hat\lambda_{-},\hat\lambda_{+}]$; the vertical dashed line locates the edge $\hat\lambda_{+}$.
Note that $W^2$ is rescaled by a factor $0.1$ to fit in the figure.\label{fig1}}
\end{figure}

We define $W^{2}(\lambda,\hat{\lambda})$ as the mean squared overlap, multiplied by $N$, between the eigenvectors of $C$ and $\hat C$ associated to, respectively, the eigenvalues $\lambda$ and $\hat{\lambda}$, namely:
\begin{equation}\label{yup2}
W^{2}(\lambda,\hat{\lambda})=\sum_{\ell,m}\frac{\left\langle (\boldsymbol{\xi}_\ell\cdot \hat{\boldsymbol{\xi}}_m)^2\, \delta(\lambda-\lambda_{\ell})\delta(\hat\lambda-\hat\lambda_m)\right\rangle}{N\,\rho(\lambda)\,\hat\rho(\hat\lambda)}.
\end{equation}

The mean squared scalar products with the top component introduced above are given by
\begin{equation}
\left<w^{2}_m\right>=\frac{1}{N}W^{2}(\gamma,\hat{\lambda}_m).
\end{equation}
We will therefore fix $\lambda=\gamma$ in the following.

$W^2$ can be computed using statistical physics approaches to random matrix theory~\cite{BS06,P07}. For $\hat\lambda \in [\hat\lambda_-(r);\hat\lambda_+(r)]$ spanning the MP bulk spectrum one has~\cite{DP11}
\begin{equation}
W^{2}(\gamma,\hat{\lambda})=\frac{\hat\gamma(\gamma)-\gamma }{\hat{\lambda} -\hat{\gamma }(\gamma)}\label{w_curve} \ .
\end{equation}  
$W^2$ in Eq.~(\ref{w_curve}) is an increasing function of $\hat
\lambda$, which diverges for $\hat\lambda=\hat{\gamma}(\gamma)$. Note that this divergence is always located outside the MP spectrum (as in both panels of Fig.~\ref{fig1}), and coincides with the MP edge $\hat \lambda_+(r)$ for the critical signal eigenvalue $\gamma=\gamma_c(r)$ only. 

It is easy to obtain the expression for the overlap between the top eigenvectors of $C$ and $\hat C$ by exploiting the standard relation for orthonormal bases, $\sum_{m}^{N} w^{2}_m=1$, or its continuous version in the infinite--$N$ limit:
\begin{equation}\label{yup}
\int_{\hat\lambda_{-}}^{\hat\lambda_{+}}W^{2}(\gamma,\hat\lambda)\hat\rho_{_{MP}}(\hat\lambda)d\hat\lambda+\langle w_1^2\rangle =1.
\end{equation}
We deduce from Eqs.~(\ref{MPdistr},\ref{w_curve},\ref{yup}) that the mean squared overlap between the top components of the population and sample covariance matrices is given by
\begin{equation}
\langle w_1^2\rangle= \left\{ \begin{array}{c} 0 \quad\qquad\qquad \gamma\leq \gamma_{c}(r),\\
 \frac{\gamma^2-\hat\gamma(\gamma)}{ \hat\gamma(\gamma) \, (\gamma- 1)} \qquad \gamma> \gamma_{c}(r)\ .\end{array}\right.\label{eq:Wtop}
\end{equation}
and is non-zero in the strong signal regime only. This result agrees with previous applications, {\em e.g.}  to signal detection~\cite{N08}, or to the inference of relevant modes in inverse problems~\cite{CMS11}.


\section{Mutual information between the empirical eigenvectors  $\hat{\boldsymbol{\xi}}_m$ and the top component  $\boldsymbol{\xi}_1$}  

As a
  consequence, close to the transition point (both from above and from
  below) the mean squared overlap $W^{2}(\gamma, \hat \lambda)$ has a strong asymmetric
  shape (see black curves in Fig.~\ref{fig1}) showing how the sample
  eigenvectors close to the right edge are highly correlated to the
  top component. This correlation is informative and can be exploited to
  infer the principal component with appropriate algorithms, as we
  show in the next section. 

To quantify this information about the component $\boldsymbol{\xi}_1$ contained in the eigenvectors $\hat{\boldsymbol{\xi}}_m$ we introduce the mutual information $I$ between those variables. $I$ is defined as the difference between the entropy of variable $\boldsymbol{\xi}_1$ and the entropy of $\boldsymbol{\xi}$ conditioned to $\{\hat{\boldsymbol{\xi}}_m\}$; it measures how much the knowledge of $\{\hat{\boldsymbol{\xi}}_m\}$ reduces the uncertainty on the estimate of $\boldsymbol{\xi}$. $I$ is non-negative and vanishes if and only if $\boldsymbol{\xi}$ and $\{\hat{\boldsymbol{\xi}}_m\}$ are independent \cite{cover}.

\begin{figure}[h]
\begin{center}
\includegraphics[width=0.7\columnwidth,clip=true]{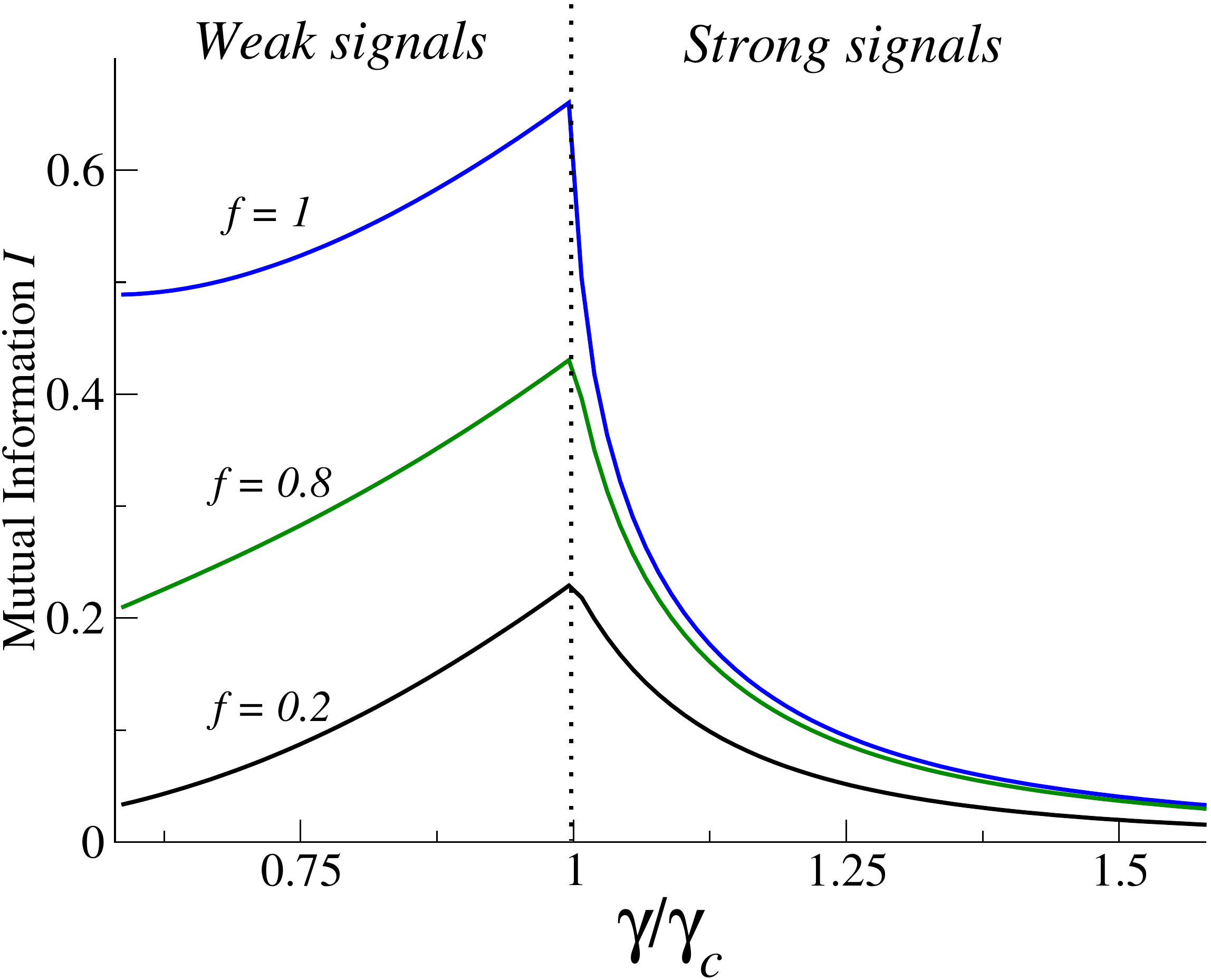}
\caption{Mutual information
  $I(\boldsymbol{\xi}_{1},\{\hat{\boldsymbol{\xi}}_m\})$,
  Eq.~(\ref{ibottom}) for the spiked covariance model
  with $r=0.5$, divided by the number $N$ of variables. Qualitatively
  similar curves are obtained when $r$ is varied. \label{fig2}  The
vertical dotted line corresponds to the transition point, below which the top empirical eigenvector is completely uncorrelated with the true (population) one, as entailed by Eq.~(\ref{eq:Wtop})}. Remarkably, the mutual information, $I$ in Eq.~(\ref{ibottom}), between an extensive number of empirical eigenvectors corresponding to lower eigenvalues and the true top components is finite positive even in this low sampling regime.
\end{center}
\end{figure}

Our goal is to understand whether the knowledge of many (of the order of $N$) very small (of the order of $1/\sqrt N$) overlaps with the empirical eigenvectors is helpful to determine the top component, {\em i.e.} if $I$ has a finite limite in the $N\to\infty$ limit. As the exact computation of $I$ is hard, we assume that the correlations between the overlaps are negligible for large $N$. Within this assumption, we calculate the mutual information $I$ between $\boldsymbol{\xi}_1$ and $\{\hat{\boldsymbol{\xi}}_m; 2\le m\le f\,N\}$, with $f\le 1$ is the fraction of retained empirical eigenvectors. The details of the calculation, based on the use of the replica approach, are given in Appendix. Within the replica symmetric framework, we obtain 
\begin{eqnarray}\label{ibottom}
\frac 1N I\big(\boldsymbol{\xi}_{1},\{\hat{\boldsymbol{\xi}}_{m}\}\big) &=& -\frac 12 \min\limits_{q, \hat q} \bigg[
 \big(q- \Omega(f)\big)\, \hat q+\textrm{log}\left(1-q\right)\nonumber\\
+ \int_{\Lambda (f)}^{\hat\lambda_+} &d\hat\lambda& \hat\rho(\hat\lambda)\  \textrm{log}\big(1+\hat q\, W^{2}(\gamma,\hat\lambda)\big)\bigg],
\end{eqnarray} 
where $\Lambda(f)$ is such that $f=\int_{\Lambda
  (f)}^{\hat\lambda_+(r)} d\hat\lambda \,\hat\rho(\hat\lambda)$ and
$\Omega(f)=\int_{\Lambda (f)}^{\hat\lambda_+(r)} d\hat\lambda
\,\hat\rho(\hat\lambda) W^2(\gamma, \hat \lambda)$. The mutual information is plotted in Fig.~\ref{fig2}
as a function of $\gamma$, and for various
values of $f$.  It is strictly positive for all values of $\gamma$. $I\big(\boldsymbol{\xi}_{1},\{\hat{\boldsymbol{\xi}}_{m}\}\big)$
reaches a maximum at the transition point $\gamma_{c}(r)$, separating the weak and strong signal regimes. Moreover it is increasing with the fraction $f$. Our calculation therefore provides clear evidence for the fact that sample eigenvectors in the bulk of the MP spectrum are informative about the principal component of the population covariance matrix. This result is remarkable especially in the case of weak signals, where the top empirical eigenvector is not informative at all. 

\section{Inference of principal component with prior knowledge}  

Based on the study of the overlaps $w_m$ above we may express the principal component $\boldsymbol{\xi}_1$ as a weighted sum of the sample eigenvectors,
\begin{equation}\label{like}
{\boldsymbol{\xi}}_{1}=\sqrt{\langle w_1^2\rangle}\;  \hat{\boldsymbol\xi}_1 + \sum_{m=2}^N\, \sigma_m \; \sqrt{\langle w_{m}^2\rangle}\;\boldsymbol{\hat{\xi}}_{m}\ ,
\end{equation}
where the $\sigma_m$'s are independent Gaussian variables of zero means, and unit variances ($m\ge 2$). Equation~(\ref{like}) implicitly defines the
likelihood of the component $\boldsymbol{\xi}_1$ given the sample eigenvectors. The rationale underlying Eq.~(\ref{like}) is that it gives back the right statistics for the
  scalar products $w_m$. In particular, the expected value (over the
      $\sigma_m$'s) of  $\boldsymbol{\xi}_{1}\cdot\boldsymbol{\hat\xi}_{m}$ vanishes, while the average value of  $(\boldsymbol{\xi}_{1}\cdot\boldsymbol{\hat\xi}_{m})^2$ coincides with 
 $ \langle w^{2}_{m}\rangle$. 
 
 In the absence of any prior
  information the average value of $\boldsymbol{\xi}_{1}$ is simply
  equal to $\sqrt{\langle w_1^2\rangle}\, \hat{\boldsymbol\xi}_1$,
  corresponding to the standard estimate widely used in
  literature. Actually this estimate discards the information contained
  in the sample eigenvectors $\boldsymbol{\hat{\xi}}_{m}$, and vanishes
  in the weak signal regime. Our purpose here is to improve over this simple estimate, by exploiting some prior information over the top component.


\begin{figure}[!h]
\includegraphics[width=0.7\columnwidth,clip=true]{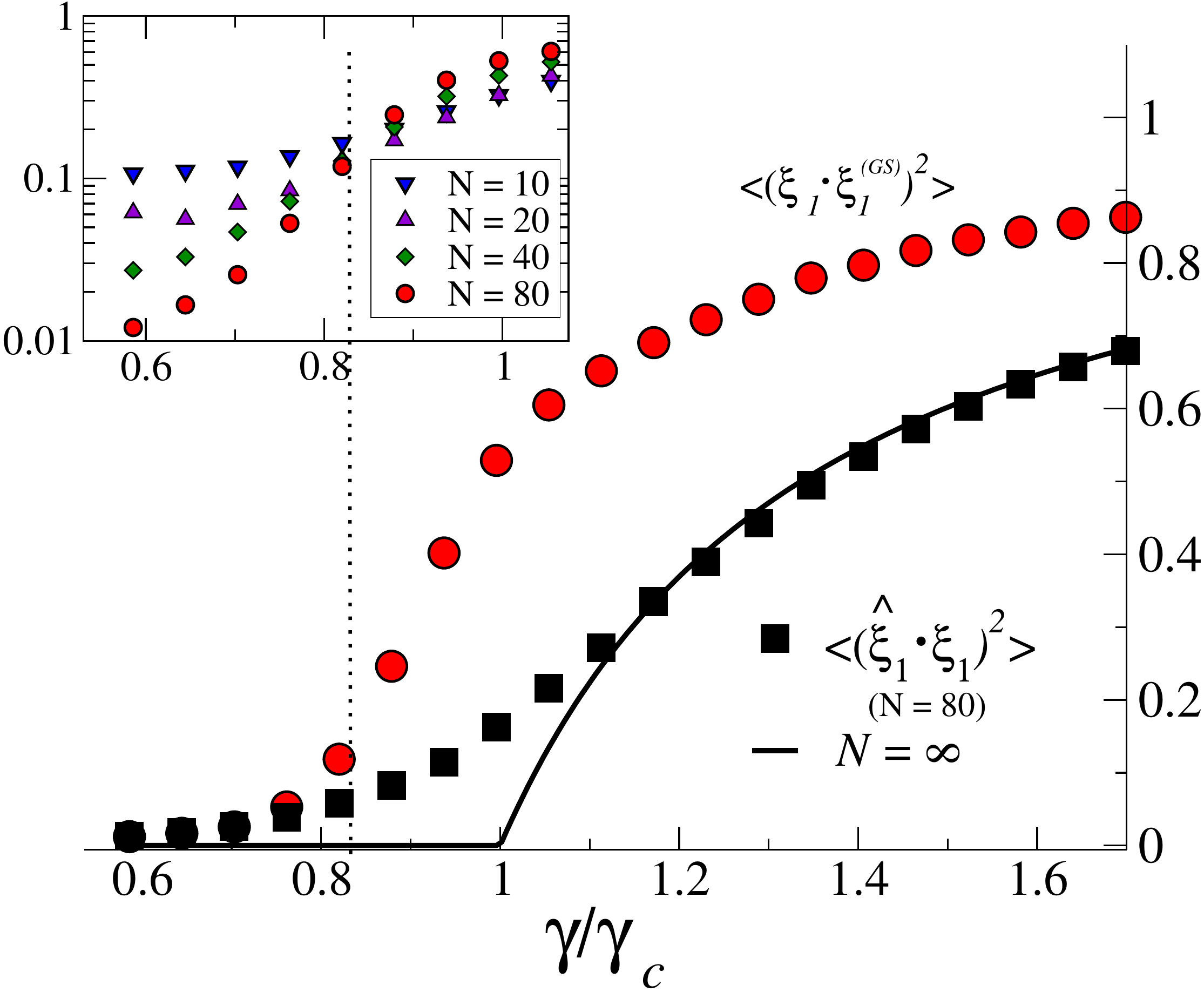}
\caption{Overlap of $\boldsymbol{\xi}_{1}\equiv (1,0,\dots,0)$
  with the ground state $\boldsymbol{\xi}^{_{(GS)}}_{1}$ of $-$IPR
  (red circles) and with the sample top component
  $\hat{\boldsymbol{\xi}}_{1}$ (black squares) as a function of
  $\gamma$ for the spiked covariance model with $r=0.5$
  ($\gamma_c\simeq 1.7071$) and $N=80$ variables. Inset: zoom of the
  region slightly below $\gamma_{c}$; overlap of
  $\boldsymbol{\xi}_{1}$ with the ground state of $-$IPR for different
  sizes $N$. All lines seem to intersect around $\gamma/\gamma_c
  \simeq 0.83$ in the poor sampling phase.}
\label{fig3}
\end{figure}

In many practical applications, indeed, prior knowledge over the principal components is available, such as the entries of those components are positive, sparse, bounded from above, etc ... A physically-sound prior knowledge we consider hereafter is the localization of  principal components, found to be important for the identification of site in contacts on the three--dimensional structure of proteins~\cite{CMW13}, or in the study of phonons in liquid crystals~\cite{C13,MS13}. Drawing our inspiration from condensed matter physics we consider the inverse participation ratio 
\begin{equation}
\textrm{IPR}(\boldsymbol{\xi}_1)=\sum_{i=1}^N[(\boldsymbol\xi_1)_i]^{4} \ ,
\label{ipr}
\end{equation}
and look for estimates of the principal component with large IPR. This prior favors vectors with large entries, {\em i.e.} non vanishing in the large--$N$ limit. More precisely the objective function to be maximized is the log-posterior distribution of $\boldsymbol\xi_1$, which sums the IPR in Eq.~(\ref{ipr}) and the log-likelihood implicitly defined by Eq.~(\ref{like}). To simplify this computational problem we consider a discrete version of Eq.~(\ref{like}), where the $\sigma_m$'s are constrained to take $\pm 1$ values. As a result we have at our disposal a pool of $2^{N-1}$ candidate components for $\boldsymbol\xi_1$ with {\em equal} log-likelihoods. We can then simply look for the binary configuration $\{\sigma_m; m\ge 2\}$ in this pool, which maximizes the IPR.


\begin{figure}[!h]
\includegraphics[width=0.8\columnwidth,clip=true]{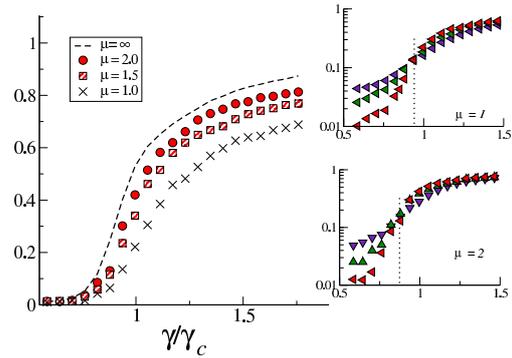}
\caption{Same simulation as in Fig.~\ref{fig3} with
  a different top eigenvector $(\boldsymbol{\xi}_{1})_{i}\propto
  \frac{1}{i^{\mu}}$, where the reconstruction is shown for
  different values of $\mu$. Remarkably the second transition is
  present also in this case (with a threshold depending on $\mu$), as
  shown in the insets for different values of $N$ ($N=20, 40, 80$).}
\label{fig3bis}
\end{figure}

To find the ground state of $-$IPR, denoted by
$\boldsymbol{\xi}^{_{(GS)}}_{1}$, we resort to simulated annealing
with a Monte Carlo scheme~\footnote{The inverse temperature $\beta$ is
  slowly incremented by steps of $\delta \beta= 5$, from $\beta= 50$
  to $\beta=200$. For each temperature we attempt 20$\,N$ Monte Carlo
  spin flips. At the end of this procedure, the configuration with
  lowest energy is retained. Average is taken over $\sim 10^{3}$
  realizations of the measurement matrix $X$.}. The overlap between
$\boldsymbol{\xi}^{_{(GS)}}_{1}$ and $\boldsymbol{\xi}_{1}$ is shown
for different sizes $N$ in Fig.~\ref{fig3}.  We find a better (higher)
value than the one corresponding to the naive estimate based on
$\hat{\boldsymbol{\xi}}_{1}$. The improvement is maximal for $\gamma$
close to $\gamma_c(r)$, that is, in the critical region separating
weak from strong signals. Remarkably, while
the naive estimate breaks down for $\gamma < \gamma_c(r)$ regime in
the large $N$ limit this does not seem to be the case for our
procedure. This result is in agreement with the
  prediction given by the Mutual Information (Eq.~(\ref{ibottom})) in Fig.~\ref{fig2}, which
  is positive even for low signals and reaches its maximum in correspondence of the transition. Therefore, the contribution of lower eigenvectors in the estimation is prominent in this region, as expected.

This scenario does not qualitatively depend on the choice of the
eigenvector $\boldsymbol\xi_1$. In particular we have studied both the
cases of a very sparse eigenvector (see Fig.~\ref{fig3}, where
$\boldsymbol\xi_1=(1, 0,\dots,0)$) and of a slow, power-law decay,
$(\boldsymbol\xi_{1})_{i}\propto \frac{1}{i^{\mu}}$ with
$\mu\ge\frac{1}{2}$. As shown in Fig.~\ref{fig3bis}, our procedure
results in a better prediction of the top eigenvector for all the
values of $\mu$ we have tested. We stress again that the  estimator in Eq.~(\ref{like}) exploits the information contained in the lower eigenmodes. A random search around $\hat{\xi_{1}}$, for  instance by considering an estimator such as $ a\,\hat{\xi_{1}}+\sqrt{1-a^{2}} \,\eta$, where $a=\sqrt{\left<w^{2}_{1}\right>}$ and $\eta$ is a
  random vector orthogonal to $\hat\xi_1$, would not produce comparable
  results, especially in the weak signal region.

\section{Transition in prior knowledge-based inference}
  
\begin{figure}[!h]
\includegraphics[width=0.9\columnwidth,clip=true]{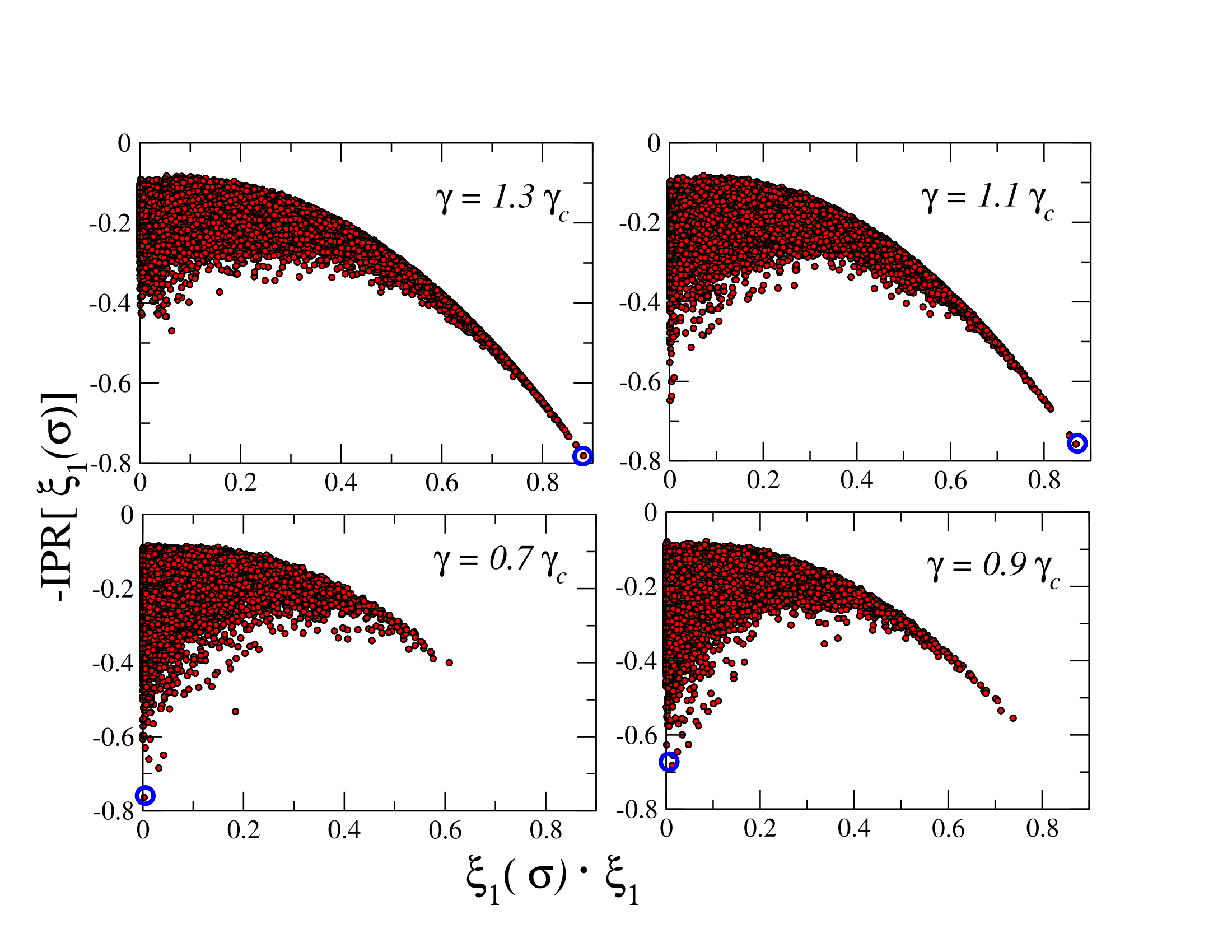}
\caption{\label{distr_stati} Typical distributions of energies of the states $\boldsymbol\sigma$ vs. the overlaps with the true eigenvector, above (top) and below (bottom panels) the transition value. The ground state is shown by the blue circle. Same model as in Fig.~\ref{fig3}, with $N=15$ and $r=0.5$. For large $N$ we expect the transition to take place at $\gamma/\gamma_c\simeq 0.83$.}
\end{figure}

As reported above, the insets of Fig.~\ref{fig3} and Fig.~\ref{fig3bis} suggest the existence of a transition point, well inside the weak signal regime, above which $\boldsymbol\xi_1$ may be approximately inferred with the help of prior knowledge, even for large system sizes. This transition bears a strong analogy with transitions taking place in the Hopfield model {\em below the critical capacity}. Indeed, the IPR in Eq.~(\ref{ipr}), once expressed in terms of the $\sigma_m$'s, may be interpreted as minus the Hamiltonian of an effective spin system, with a mixture of $k$-body interactions, where $k$ ranges from 1 to 4. The interactions 
\begin{equation}
J_{m_1,m_2,m_3,m_4}= \sum_{i=1}^N \prod_{\ell=1}^4 \sqrt{\langle w_{m_\ell}^2\rangle}\, \hat{\xi}_{i,m_\ell}
\end{equation}
are non-linear combinations of the eigenmodes components, and may have positive or negative signs. This spin-glass Hamiltonian is strongly reminiscent of  the Hopfield model \cite{hop82}. Each entry $(\boldsymbol\xi_1)_i$, $i=1,...,N$, of the principal component may be interpreted as the `magnetization' $M_i$ of the spin configuration $\boldsymbol\sigma$ along the `pattern' $i$, whose $m^{th}$ entry is $\sqrt{\langle w_m^2\rangle}(\hat{\boldsymbol\xi}_m)_i$, or, equivalently,
\begin{equation}
M_i (\boldsymbol\sigma) = \sum_{m=1}^N \sigma_m \; \sqrt{\langle w_m^2\rangle}(\hat{\boldsymbol\xi}_m)_i
\end{equation}
with $\sigma_1\equiv 1$. Note that our Hamiltonian,
\begin{equation}
-\textrm{IPR}\big(\boldsymbol{\xi}_1 (\boldsymbol\sigma)\big)=\sum_{i=1}^N M_i(\boldsymbol\sigma)^{4} \ ,
\label{ipr4}
\end{equation}
 is however quartic, and not quadratic in the magnetizations. Thanks to this analogy, the transition observed here can be interpreted in terms of the phase diagram of the Hopfield model \cite{amit85}: At low temperatures and intermediate loads (comprised between $\simeq 0.05$ and the critical capacity $\simeq 0.14$) the patterns to be stored are local minima of the Hopfield Hamiltonian, and are uncorrelated with the ground state. This scenario holds in our case too. We show in Fig.~\ref{distr_stati} the distribution of the energies vs. the overlap, obtained through exhaustive searches of the configuration space for small sizes $N<25$. Below the transition point, the ground-state vector $\boldsymbol{\xi}^{_{(GS)}}_{1}$ is clearly not aligned along $\boldsymbol\xi_1$. 

\section{Conclusions} 
As described by random matrix theory, the overlap between lower
empirical eigenvectors and the top true one is very small, of the order
of $N^{-1/2}$ and vanishes in the infinite $N$ limit. In spite of this,
in this paper we have shown how an extensive number of sample
eigenvectors with low eigenvalues is strongly informative about the
population principal components,  by presenting a calculation of mutual information for the spike covariance model.

Based on this result, we have introduced a general procedure to exploit that
information in the presence of prior knowledge, by mapping
the inference problem onto the search for the the ground state of a
spin-glass-like Hamiltonian encoding the prior. We have shown the
efficiency of the approach when one knows {\em a priori} that top
components have large entries, which considerably improves the standard
inference and allows us to recover the component in the weak signal
region, where naive inference fails. This finding agrees with recent
results on non-negative PCA~\cite{MR14}.

It would be interesting to understand how efficient is our procedure for other 
priors, or in cases where the value of the overlap distribution is unknown and
eigenvalue-cleaning techniques~\cite{DP11,BABP15}) must be used for
its estimation. A limitation of our approach is the use of discrete (binary) variables $\sigma_m$ in Eq.~(\ref{like}). In a forthcoming publication we plan to study more refined algorithms by considering continuous variables instead, in order to
test the generality of the (second) transition found in the weak
signal regime. 

We stress that our approach could also be extended to
infer more than one principal components. While the case of a finite
number of separated eigenvalues (multiple-spiked covariance model) is
straightforward, it would be interesting to consider
$O(N)$-dimensional degenerate subspaces, as in~\cite{M08}.



\acknowledgments
 We are very grateful to S. Cocco and G. Robichon for their participation to the initial stage of this work. We thank
M.P. Brenner, M. Huntley, P. Vivo for discussions, and P. Vivo for a careful reading of the manuscript. This work was partly funded by the Agence Nationale de la
Recherche Coevstat project (ANR-13-BS04-0012-01) and the CNRS-InphyNiTi project INFERNEURO.

\section{Appendix: replica calculation of the mutual information}

Here, we present the derivation of the mutual information Eq.~(\ref{ibottom}). We assume that the components of $\hat{\boldsymbol\xi}_m$ are independent and identically Gaussianly distributed, with zero means and unit variances, and that the dot products $w_m$ are also independent normal variables with zero means and variances $W^2_m/N$. The index $m$ runs from 2 to $M\equiv f\,N$, where $f$ is the fraction of eigenvectors retained. Equation (\ref{like}) thus fully defines the joint distribution of the eigenvectors,  $P[\boldsymbol\xi_1, \{\hat{\boldsymbol\xi}_m\}] $. We define $Z(n) = \int d\boldsymbol\xi_1\prod_{m} d\hat{\boldsymbol\xi}_m \; P[\boldsymbol\xi_1, \{\hat{\boldsymbol\xi}_m\}] ^n$.
$Z(1)$ is obviously equal to unity by normalisation of $P$, and the mutual information $I$ is simply related to the derivative of $Z$ in $n=1$, see below. Along the lines of the replica method, we will first consider that $n$ is an integer, and will next perform an analytic continuation to real-valued $n$ on the outcome. After integrating out the eigenvector components and within the replica symmetric hypothesis, we obtain that $F_n\equiv\frac 1N \log Z(n)$ is given, up to some irrelevant additive constant, by the saddle-point value of 
\begin{eqnarray}
F_n &=& -\frac {n(n-1)}2 q\, s - \frac n2 \,\tilde q\, \tilde s  \\
&-& \frac 12 \log\left[ n \left(1-\sum_m W_m^2+\frac{\tilde q -q}n\right)^{n-1}\right] \nonumber \\
&-& \frac{n}{2N}  \sum_{m}  \log\left( 1+W_m^2 (s-\tilde s)\right)\nonumber  \\
&-&\frac 1{2N}\sum _m  \log\left( 1-\frac{n \,s\, W_m^2}{1+W_m^2 (s-\tilde s)}\right)  \bigg]\nonumber
\end{eqnarray}
over its four arguments $\tilde q,q,\tilde s,s$. The parameters $\tilde q,q$ are equal to, respectively, $\sum_m \overline{ \langle w_m^2\rangle}, \sum_m \overline{\langle w_m\rangle^2}$,   where the overbears denotes the averages over the eigenvectors and the brackets denotes the averages over the Gaussian measure over the overlaps at fixed eigenvectors; $\tilde s, s$ are the conjugated Lagrange parameters.

A straightforward calculation shows that the mutual information $I$ in Eq.~(\ref{ibottom}) is given by
$\frac 1N I\big(\boldsymbol{\xi}_{1},\{\hat{\boldsymbol{\xi}}_{m}\}\big) = -\left.\frac{\partial F_n}{\partial n} \right|_{n=1}$. 
We are therefore left with the resolution of the saddle-point equations over $\tilde q, q,\tilde s$, and $s$. First, let us remark that $F_1$ depends on $\tilde q$ and $\tilde s$ only: $F_1= -\frac {\tilde s\, \tilde q}2 - \frac 1{2N}\sum_m\log\left( 1-\tilde s\; W_m^2 \right)$.
The values of the order parameters at the saddle point are thus $\tilde q=\frac 1N \sum_m W_m^2$, $\tilde s=0$.  We next consider $F_{n=1+\epsilon}=F_1 - \epsilon I+O(\epsilon^2)$, where
\begin{eqnarray}
I&=& \frac {s\, q}2  +\frac 1{2N}\sum_m\log\left( 1+s\; W_m^2 \right) \nonumber \\
&-& \frac {s}{2N} \sum_m W_m^2+\frac 12 \log(1-q) \ .
\end{eqnarray}
To the lowest order in $\epsilon$, the values of $\tilde q$ and $\tilde s$ are unchanged with respect to the case $n=1$. The saddle-point equations for $q$ and $s$ give 
$s =\frac 1{1-q},  q =\frac sN \sum_m \frac{W_m^4}{1+s\, W_m^2}$.
The corresponding expression for $I$ is given in Eq.~(\ref{ibottom}).

\end{document}